%% file: main.tex
\documentclass[aps,prfluids,12pt,tightenlines,onecolumn,amsmath,amssymb]{revtex4-2}
\usepackage{mathptmx}
\usepackage{hyperref}
\hypersetup{colorlinks=true,linkcolor=blue,urlcolor=blue,citecolor=blue}
\usepackage{newtxmath} % emulates mtpro2 font
\input{babel-mod} % errors with babel because of language aliases
\usepackage{graphicx} % Include figure files
\usepackage{dcolumn} % Align table columns on decimal point
\usepackage{siunitx} % SI units
\usepackage[caption=false, justification=justified]{subfig} % Subfigures - do not use subcaption package, incompatible with RevTex4.2
\usepackage{chngcntr}
\input{caption-mod}
\usepackage[export]{adjustbox}

\usepackage{color}
\usepackage[normalem]{ulem}

\newcommand{\BLUE}[1]{{\color{black}#1}}
\definecolor{darkgreen}{rgb}{0.0, 0.5, 0.0}

\begin{document}
\newcommand{\cselab}{
	a. Computational Science and Engineering Laboratory, ETH Z\"{u}rich, CH-8092, Switzerland.}

\newcommand{\harvard}{
	b. School of Engineering and Applied Sciences, Harvard University,
	Cambridge, MA 02138, USA.}
\renewcommand{\appendixname}{APPENDIX}
\renewcommand\figurename{FIG.} % Figure captions as per PRFluids
\title{Learning swimming escape patterns {\BLUE{for  larval fish}} under energy constraints}
\author{Ioannis Mandralis}
\author{Pascal Weber}
\author{Guido Novati}
\author{Petros Koumoutsakos}
\thanks{petros@seas.harvard.edu}
%	\altaffiliation[Also at ]{\harvard}
\affiliation{\cselab,\harvard}  
\date{\today}
% \keywords{reinforcement learning, fluid mechanics}

\begin{abstract}
Swimming organisms can escape their predators by creating and harnessing unsteady flow fields through their body motions. Stochastic optimization and flow simulations have identified escape patterns that are consistent with those observed in natural larval swimmers. However, these patterns have been limited by the specification of a particular cost function and depend on a prescribed  functional form of the body motion. Here, we deploy reinforcement learning to discover swimmer escape patterns for larval fish under  energy constraints. The identified patterns include the C-start mechanism, in addition to more energetically efficient escapes. We find that maximizing distance with limited energy requires swimming via short bursts of accelerating motion interlinked with phases of gliding. The present, data efficient, reinforcement learning algorithm results in an array of patterns that reveal practical flow optimization principles for efficient swimming and the methodology can be transferred to the control of aquatic robotic devices operating under energy constraints.
\end{abstract}

\maketitle

\section{Introduction}
Aquatic organisms involved in predator-prey interactions perform impressive feats of fluid manipulation to enhance their chances of survival \cite{triantafyllou_survival_2012, peng_transport_2009, nair_fish_2017, mchenry_pursuit_2019, borla_prey_2002, soto_when_2015, budick_locomotor_2000, colin_stealth_2010}. Since early studies where prey fish were reported to rapidly accelerate from rest by bending into a C-shape and unfurling their tail \cite{gray_directional_1933, webb_acceleration_1975, weihs_mechanism_1973, domenici_kinematics_1997}, impulsive locomotion patterns have been the subject of intense investigation. Studying escape strategies of prey fish has led to the discovery of sensing mechanisms \cite{bollmann_zebrafish_2019, carrillo_canal_2019, stewart_prey_2014}, dedicated neural circuits \cite{bianco_visuomotor_2015, bianco_prey_2011, dunn_brain-wide_2016, dunn_neural_2016}, and bio-mechanic principles \cite{jayne_red_1993, schwalbe_red_2019}. From the perspective of hydrodynamics, several studies have sought to understand the C-start escape response and how it imparts momentum to the surrounding fluid \cite{witt_hydrodynamics_2015, epps_impulse_2007, tytell_hydrodynamics_2008, li_escape_2014, borazjani_functional_2013, borazjani_hydrodynamics_2012}. 

Despite the large volume of literature on the C-start escape response, experiments and observations indicate that swimming escapes can take a variety of forms. For example, after the initial burst from rest, many fish are seen coasting instead of swimming continuously \cite{weihs_mechanism_1973, kramer_behavioral_2001, chadwell_median_2012}. Furthermore, theoretical and experimental studies have suggested that intermittent swimming styles, termed burst-coast swimming, can be more efficient than continuous swimming when maximizing distance given a fixed amount of energy  \cite{weihs_energetic_1974, weihs_energetic_1980, videler_energetic_1982, wu_kinematics_2007}. This raises the question of when and why different swimming escape patterns are employed in nature, and which biophysical cost functions they optimize.

This question has been investigated by using reverse engineering methodologies to identify links between biophysical cost functions and resulting swimming patterns. For example, fast and efficient swimming motions \cite{kern_simulations_2006, Kern2008} or the C-start escape response \cite{gazzola_c-start_2012} have been reverse engineered based on the appropriate objective functions. In particular, when reverse engineering the C-start escape response, the parameters of a \textit{pre-defined} motion-sequence were optimized to maximize the escape distance. While the identified escape pattern was consistent with larval fish escapes observed in nature, the reverse engineering method employed had inherent limitations: it required an a-priori defined objective function and a specification of the functional form for the two stages of the observed escape patterns. As a result, the full space of swimming escapes remained unexplored since reverse engineered escape patterns could only reside within the predefined design-space and depended strongly on the underlying parametric assumptions.

In recent years, reinforcement learning (RL), an alternative framework for \textit{learning} behavior based on objective functions, has emerged. RL has found application for a variety of problems related to fluid mechanics, including learning natural swimming and flying behaviors \cite{gazzola_reinforcement_2014, reddy_learning_2016, gazzola_learning_2016, novati_synchronisation_2017, colabrese_flow_2017-1, verma_efficient_2018,reddy_glider_2018,novati_controlled_2019, novati_deep-reinforcement-learning_2018} and optimizing and controlling engineering flow systems \cite{gueniat_statistical_2016, rabault_artificial_2019, brunton_machine_2020,fan_reinforcement_2020}. In the RL framework, decision making processes are viewed as multistage optimizations instead of one-shot optimizations that require rigidly pre-defined motion parameters. Due to this flexibility, RL circumvents many of the limitations of classical reverse engineering and offers a novel way through which to explore the space of swimming escape patterns. 

In this paper, we introduce reinforcement learning (RL) to the study of escape responses. In particular, we employ the RL framework to understand the links between objective functions and swimming escape patterns for larval fish. The fish escape is formulated as an incremental process where a swimming agent receives information about the flow field and learns to maximize its cumulative reward autonomously. By endowing the swimming agent with limited energy and rewarding the escape distance, we find that burst-coast swimming escapes, consisting of rapid body accelerations through C-bends followed by powerless gliding, maximize escape distance when the available energy is limited, in alignment with theoretical predictions \cite{weihs_energetic_1974, weihs_energetic_1980, videler_energetic_1982}. Furthermore, we find that the RL algorithm is able to produce a wide array of different escape patterns, according to the amount of energy available, due to its inherent generalization capability. This  ``kaleidoscope" of escape patterns  sheds light on key mechanisms which are responsible for rapid propulsion in fluids and evidences the fundamental advantage of using RL to provide links between objective functions and biological behavior.

%These results are compared to those obtained via reverse engineering, evidencing the fundamental advantages of employing the RL framework to provide links between objective functions and biological behavior. 

% The ability of the RL swimmer to adapt its strategy according to the amount of energy available, is a testament to the generalization capability of the algorithm. 

\section{Swimmer Model and Kinematics}
The geometry of the artificial swimmer, displayed in Figure \ref{img:system-overview}(a), is modelled after a 5 days-post-fertilization zebrafish (see Appendix for details). The swimmer propels itself by modifying its instantaneous mid-line curvature $\kappa(s,t)\in\mathbb{R}$, a quantity which imitates muscle contractions in natural anguilliform swimmers~\cite{kern_simulations_2006}. The midline curvature $\kappa(s,t)$, displayed in equation \eqref{eq:midline-curvature}, is further decomposed into a baseline component $B(s,t)\in\mathbb{R}$ and an undulatory component $K(s,t)\in\mathbb{R}$. This allows the swimmer to bend unilaterally and to undulate sinusoidally. 
\begin{equation}
\begin{split}
    \kappa(s,t) &=B(s,t)+ K(s,t) \sin\left[2\pi \left(t/T_{\textrm{prop}} - s\tau_L(t)/L\right)+ \phi(t)\right]\,.
\end{split}
\label{eq:midline-curvature}
\end{equation}
The baseline curvature $B(s,t)$ and the undulatory curvature $K(s,t)$ are modeled as natural cubic splines defined by six control points on the swimmer mid-line (see Figure \ref{img:system-overview} (a)). To imitate the stiff head, neck, and tail of larval zebrafish, the curvature is set to zero at $s_1=0, s_2=0.2 L, s_6=L$, leaving three free control points at which the curvature can be controlled. The midline curvature is parametrized by $T_{\textrm{prop}}\in\mathbb{R}$, the undulatory swimming period, $L$, the swimmer length, $\tau_L(t)$, the delay of the tail, and $\phi(t)$, the phase of the sinusoid. Given the curvature $\kappa(s,t)$, the midline coordinates of the swimmer are retrieved by solving the Frenet-Serret formulas~\cite{Kern2008}. As per in vivo observations of 5 days post fertilization zebrafish~\cite{muller_flow_2008}, the swimmer length is set to $L=4.4\ \textrm{mm}$ and the propulsive swimming period to $T_{\textrm{prop}}=44\ \textrm{ms}$. The resulting swimming Reynolds number, defined as $\operatorname{Re} = \frac{L^2}{T_{\textrm{prop}}\nu}$, where $\nu$ is the kinematic viscosity of water, is chosen as $\operatorname{Re}=550$. This places the swimmer in the intermediate flow regime where both viscous and inertial forces have important effects. The Navier-Stokes equations are solved by performing a direct numerical simulation on a uniform grid (see Appendix A for details).

\begin{figure}[t]
  \centering
  \includegraphics[width=0.9\textwidth]{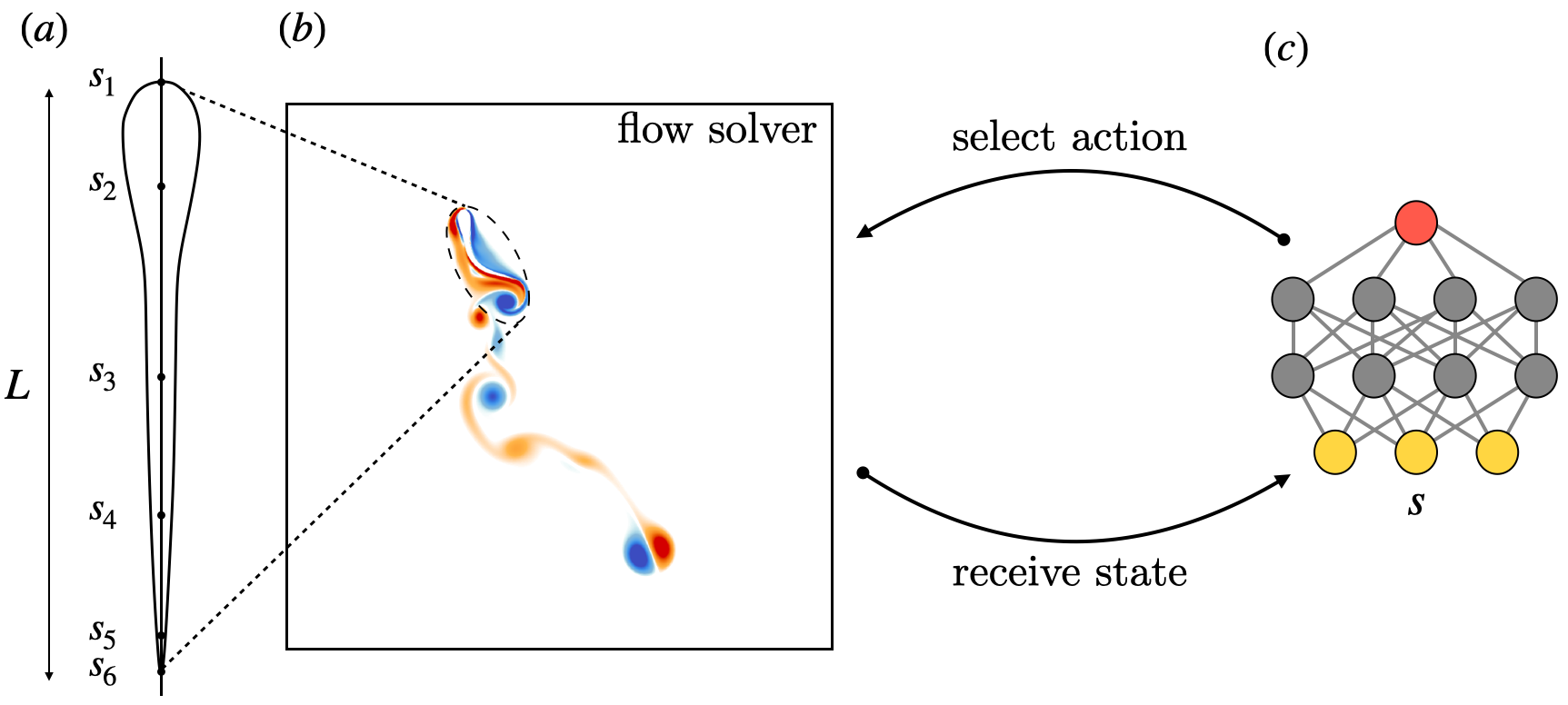}
  \caption{Interaction between flow-solver and reinforcement learning. (a) Geometrical model of a 5 days post-fertilization zebrafish larva~\cite{gazzola_c-start_2012}. The six curvature control points are indicated from top to bottom. (b) Simulation environment. The swimmer is placed inside a square domain and is simulated using the numerical flow solver described in Appendix A. (c) This is coupled with a deep reinforcement learning algorithm, that receives a state from the flow solver and sends back an action, deciding how the swimmer should act within the simulation.}
  \label{img:system-overview}
\end{figure}

\section{Reinforcement Learning for Swimming Escapes}
\label{sec:rl}
In the RL framework, an agent learns to earn rewards through trial-and-error interaction with its environment. The agent chooses an action $\boldsymbol{a}_k\in\mathcal{A}$ at discrete time instances $k\in\mathbb{N}$ by sampling a stochastic control policy $\boldsymbol{\pi}(\boldsymbol{a}|\boldsymbol{s}_k)$ that is conditioned on its current state $\boldsymbol{s}_k\in\mathcal{S}$, i.e. $\boldsymbol{a}_k \sim \boldsymbol{\pi}(\,\cdot\,|\boldsymbol{s}_k)$. Given the action, the environment transitions to a new state determined by the dynamics function $D$, i.e $\boldsymbol{s}_{k+1} \sim D(\,\cdot\,|\boldsymbol{a}_k, \boldsymbol{s}_{k})$. Upon transition, the agent receives a reward signal $r_{k+1}\in\mathbb{R}$. The goal is to learn the optimal control policy $\boldsymbol{\pi}^\star(\boldsymbol{a}|\boldsymbol{s})$ which maximizes the action-value function $Q\left(\boldsymbol{s}, \boldsymbol{a}\right)\in\mathbb{R}$, defined as the expected long-term cumulative reward when starting from state $\boldsymbol{s}$ and taking action $\boldsymbol{a}$
\begin{equation}
    Q(\boldsymbol{s}, \boldsymbol{a}) =\mathbb{E}_{\substack{\boldsymbol{a}_k\sim\boldsymbol{\pi} \\ \boldsymbol{s}_{k+1} \sim D }}\left[\sum_{k=0}^{N}\gamma^k r_{k}\Bigg|\boldsymbol{s}_0 = \boldsymbol{s}, \boldsymbol{a}_0 = \boldsymbol{a}\right]\,.
\end{equation}
Here, $\gamma\in[0,1)$ is the discounting factor which quantifies the trade-off between immediate and future rewards. We use a state of the art and  data efficient RL algorithm (Remember and Forget Experience Replay) to identify the optimal policy~\cite{novati_remember_2019} (see Appendix B for details).

To model the escape behavior of a prey, the swimmer is trained to maximize the distance away from its initial position after a fixed time. In addition, the work done by the swimmer on the fluid is limited based on an escape energy budget $E_0$. This imitates how the  muscle fibers used during fish escapes can be fatigued by lactic acid build-up as glycogen stores are depleted \cite{videler_energetic_1982}. We normalize these budgets by $E_0$ which is the energy expended during the C-start escape sequence reported by Gazzola et. al. \cite{gazzola_c-start_2012} that closely matches larval zebrafish escapes observed in nature ($E_0=14.04, d_0=1.15 L$, see Appendix B for details on computation and non-dimensionalization). 

Each escape episode proceeds by first sampling an energy budget uniformly in a range around $E_0$ ($[\frac{1}{3}E_0, 3 E_0]$), allowing the swimmer to interact with the fluid for a fixed time period, and finally rewarding the overall distance travelled ($r=d$). If the energy budget is depleted before the allocated escape time is surpassed, the episode terminates prematurely (see Appendix B for training details). During each escape, the swimmer controls its body through a nine dimensional action vector that schedules changes in the shape of the curvature as well as the delay and phase of the sinusoidal motion. The swimmer senses the environment through a set of states (current distance and polar angle from starting position, body orientation, mean forward and angular velocity, as well as the remaining energy budget) and receives as reward its distance from the start, at the end of each episode.

\begin{figure}[hbtp]
    \centering
    \includegraphics[width=1\textwidth]{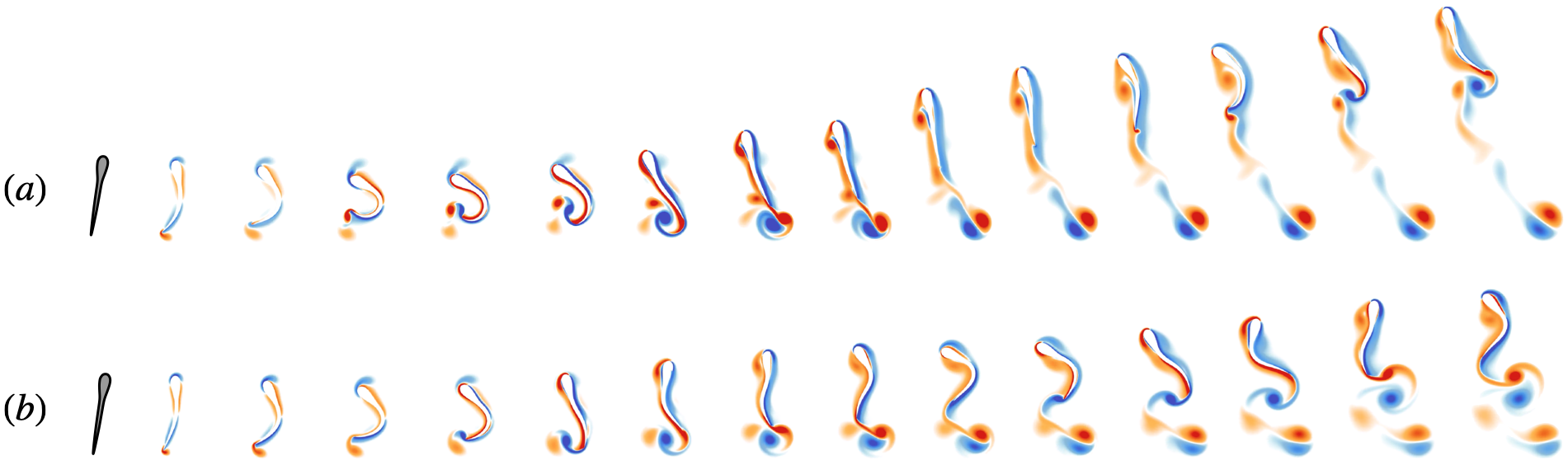}
    \caption{Temporal evolution of the vorticity field for the optimized and learned escape patterns. (a) RL burst-coast escape pattern obtained by RL swimmer using energy budget $E_0$. (b) C-start escape pattern obtained by simulating the optimized parameter set reported in \cite{gazzola_c-start_2012}. Orange regions represent positive flow vorticity and blue regions represent negative vorticity. }
    \label{img:escapes}
\end{figure}

\section{Learned vs Optimized Escapes}
\label{sec:learned-vs-optimized}
The vorticity field generated by the trained RL swimmer when escaping with energy budget $E_0$ is displayed in Figure \ref{img:escapes}(a). The swimmer initially forms a C-bend and subsequently unfurls its tail, propelling a counter-rotating vortex pair opposite to the direction of its forward motion and coasting. Only when its speed drops significantly does the learned swimmer undulate its body further in an attempt to extend its forward motion. This learned motion sequence is termed the \textit{burst-coast} escape pattern. In the following, we compare the burst-coast escape pattern obtained using RL to the \textit{C-start} escape pattern which was found by optimizing a pre-defined motion sequence in \cite{gazzola_c-start_2012}.

The vorticity fields of the two escape sequences are displayed in Figure \ref{img:escapes}(b). Figure \ref{img:escapes}(b) indicates that, although the RL swimmer coasts after the initial burst instead of swimming continuously, the C-bend starting pattern is remarkably close to that of the C-start, without having been enforced through pre-defined motion parameters. The two escape patterns differ in that the C-start has a continuous swimming phase which propels two counter-rotating vortex dipoles away from the swimmer body, whilst the burst-coast escape pattern propels only one counter-rotating vortex dipole which is created in the initial burst.

\begin{figure}[htbp]
\centering
\begin{minipage}[t]{0.15\textwidth}

\end{minipage}\hfill
\begin{minipage}[t]{0.01\textwidth}
(a)
\end{minipage}
\begin{minipage}[t]{0.50\textwidth}
\includegraphics[width=0.8\textwidth, valign=t]{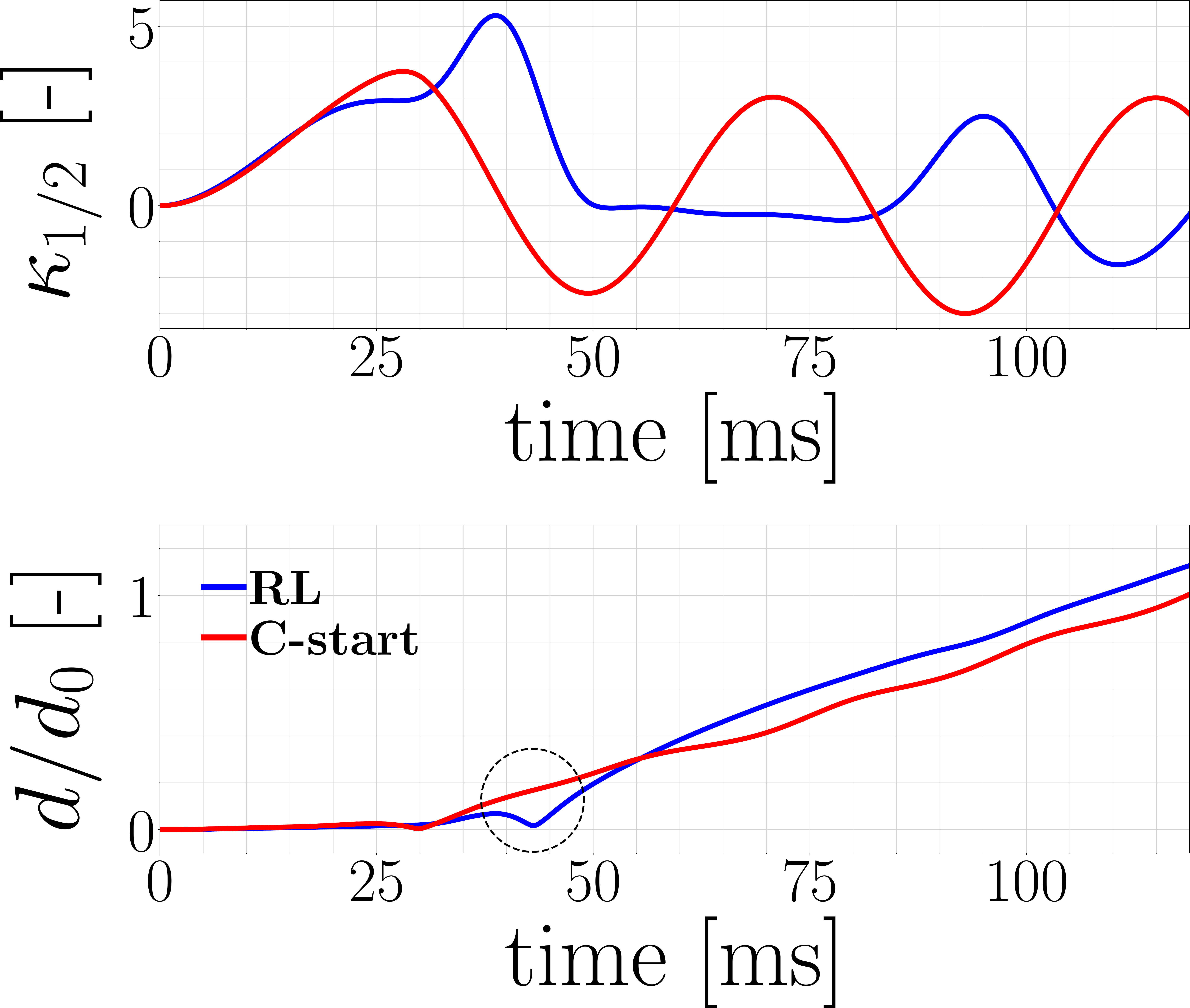}
\end{minipage}
\begin{minipage}[t]{0.03\textwidth}
(b)
\end{minipage}
\begin{minipage}[t]{0.22\textwidth}
\includegraphics[width=0.8\textwidth, valign=t]{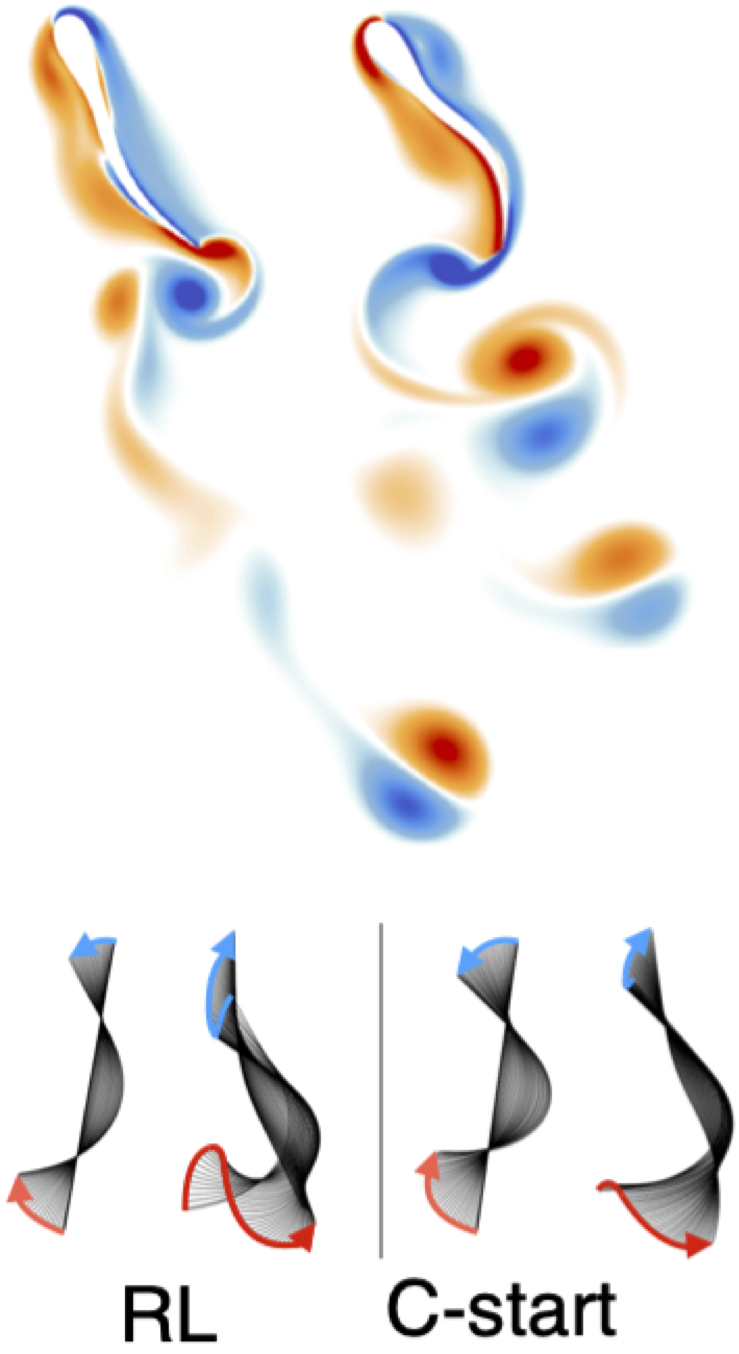}
\end{minipage}\hfill
\begin{minipage}[t]{0.15\textwidth}
\end{minipage}%
\caption{Comparison of the equal-energy RL burst-coast and C-start escape patterns. (a) Swimmer midpoint curvature $\kappa_{1/2}$ and normalized distance (in terms of swimmer length) as a function of time. The energy expended by the RL escape (blue) and C-start escape (red) is in both cases equal to $E_0$. (b) Vorticity fields and midline profiles for the RL escape (left) and the  C-start escape (right). For both escapes, the midline of the swimmer is plotted over time within two intervals: the preparatory interval ($t\le30\si{\milli\second}$) and the propulsive interval ($30\si{\milli\second}\le t \le 50\si{\milli\second}$).}
\label{img:rl-compared-to-c-start}
\end{figure}

Moreover, the burst-coast strategy results in a greater escape distance than the C-start while using an equal amount of energy, as shown in Figure~\ref{img:rl-compared-to-c-start}(a). Since both escapes consume equal energy and conform to natural mechanic/geometric constraints (see Appendix A), they are both feasible in a hypothetical predator-prey encounter. In fact, the use of both intermediate swimming or coasting phases are observed for various fish species in predator escapes~\cite{weihs_mechanism_1973, kramer_behavioral_2001, chadwell_median_2012}. In which situations is it advantageous to employ the C-start or the burst-coast escape patterns ?

We found that, at the characteristic swimming Reynolds number for larval zebrafish, $\operatorname{Re}=550$~\cite{gazzola_c-start_2012}, the C-start sets the swimmer into motion faster than the burst coast pattern (circled region Figure~\ref{img:rl-compared-to-c-start}(a)). In contrast, the RL swimmer uses a burst-coast pattern which more efficiently solves the task we set out for it - instead of propelling itself forward quickly, it delays the onset of its motion in order to form a more pronounced C-shape, evidenced by the higher peak in midpoint curvature in Figure~\ref{img:rl-compared-to-c-start}(a), and achieves a greater overall distance using the same amount of energy. This underscores the energetic advantage of the burst-coast escape pattern as compared to continuous swimming (see Section~\ref{sec:reynolds}).

\section{Influence of Reynolds number on escape patterns}\label{sec:reynolds}
\begin{figure}[htbp]
\begin{minipage}[t]{0.02\textwidth}
(a)
\end{minipage}
\begin{minipage}[t]{0.40\textwidth}
\includegraphics[width=\textwidth, valign=t]{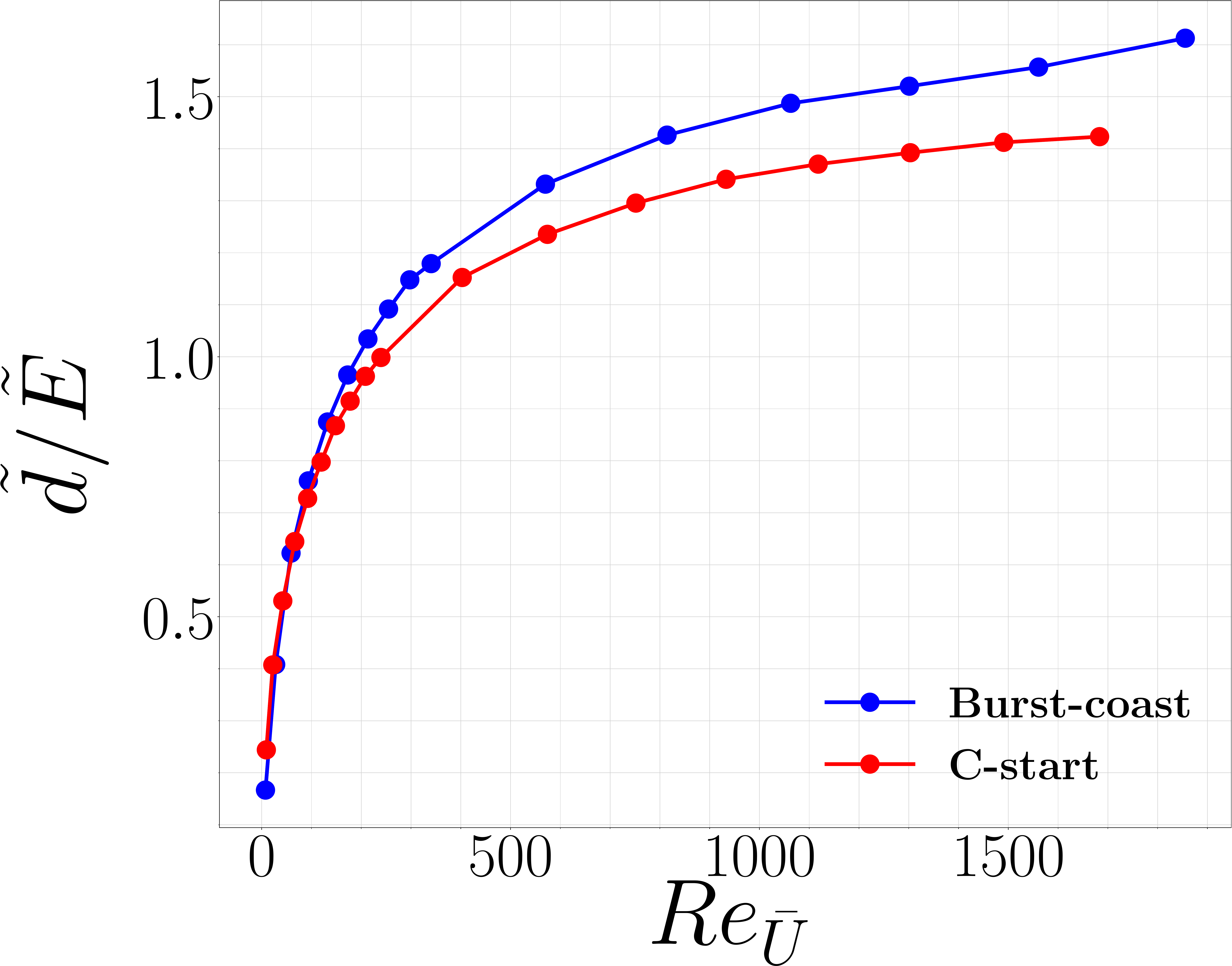}
\end{minipage}
\begin{minipage}[t]{0.04\textwidth}
\end{minipage}
\begin{minipage}[t]{0.03\textwidth}
(b)
\end{minipage}
\begin{minipage}[t]{0.40\textwidth}
\includegraphics[width=\textwidth, valign=t]{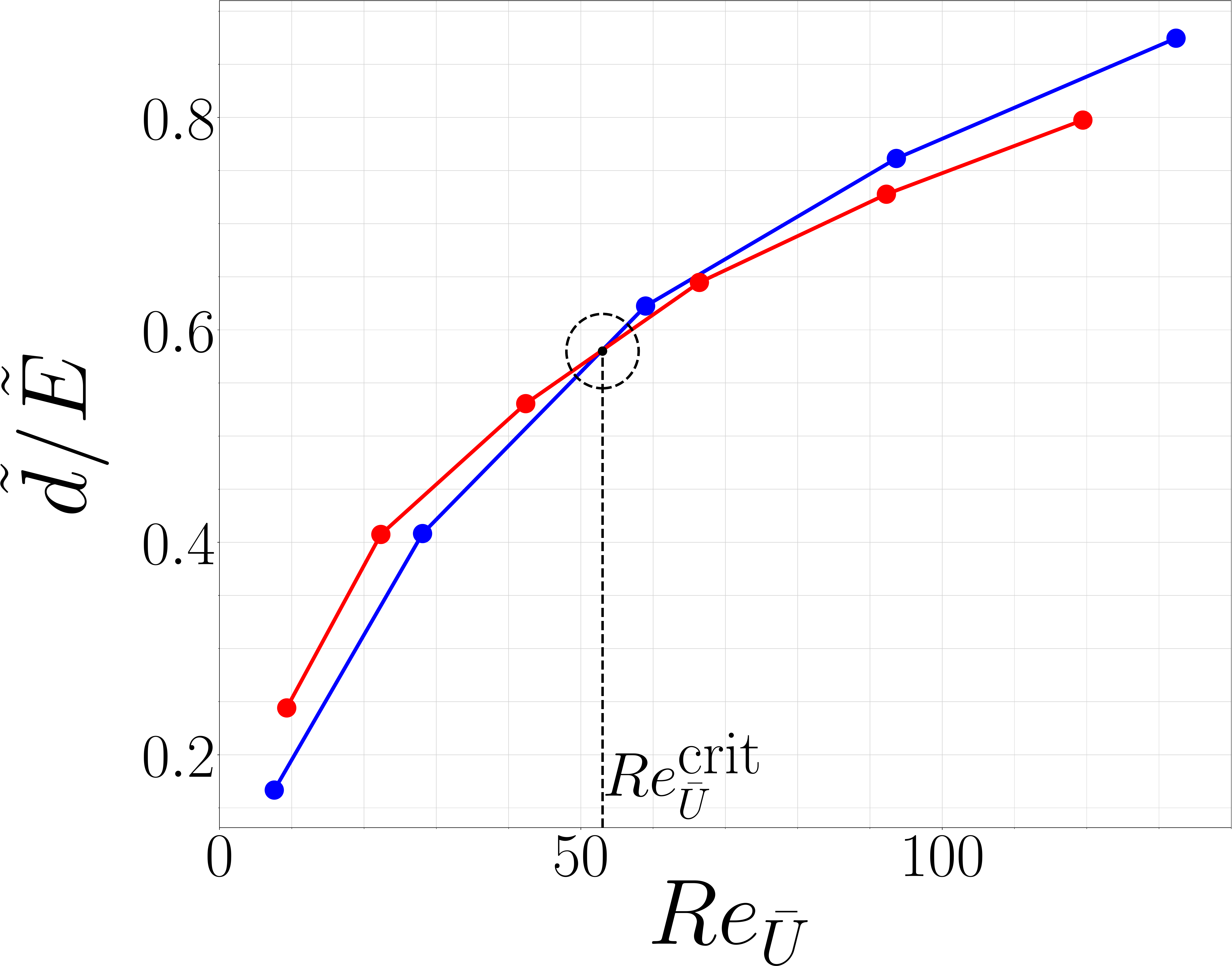}
\begin{minipage}[t]{0.20\textwidth}
\end{minipage}\hfill
\end{minipage}%
\caption{Influence of Reynolds number on escape efficiency. (a) The normalized distance per unit energy $\tilde{d}/\tilde{E}$ is plotted as a function of the mean Reynolds number $\operatorname{Re}_{\bar{U}}$. $\tilde{d}=d/d_0$ and $\tilde{E}=E/E_0$. In blue: the burst coast escape sequence recorded by evaluating the RL control policy with energy budget $E_0$ at $\nu_0$, simulated for different viscosities. In red: the C-start escape sequence obtained by optimizing at $\nu_0$ \cite{gazzola_c-start_2012}, simulated at different viscosities. (b) A zoomed in segment of Figure \ref{img:reynolds}(a) on the range $Re_{\bar{U}}\in[0,150]$. Displays the critical Reynolds number $Re_{\bar{U}}^{\textrm{crit}}$ at which the burst coast sequence becomes more efficient than the C-start.}
\label{img:reynolds}
\end{figure}

We recorded the motion sequence of the burst-coast RL escape with energy budget $E_0$ at the original viscosity $\nu_0$ (corresponding to Reynolds number $\operatorname{Re}_0 = 550$), and simulated it along with the C-start for viscosities in the range $\nu \in [0.1\nu_0, 5\nu_0]$. Based on the average speed during each escape the  $\operatorname{Re}_{\bar{U}} = \frac{\bar{U}L}{\nu}$ was computed. The normalized distance travelled per unit energy consumed $\tilde{d}/\tilde{E}$ is plotted for each escape against the mean Reynolds number $\operatorname{Re}_{\bar{U}}$ in Figure \ref{img:reynolds}. This uncovers two distinct regimes: the low Reynolds number regime $\operatorname{Re}_{\bar{U}} \lesssim \operatorname{Re}_{\bar{U}}^{\textrm{crit}} = 50$, and the high Reynolds number regime $\operatorname{Re}_{\bar{U}} \gtrsim \operatorname{Re}_{\bar{U}}^{\textrm{crit}} = 50$. In the high Reynolds number regime the discrepancy between the burst-coast escape sequence and the C-start increases. This finding is consistent with the theoretical considerations of Weihs \cite{weihs_energetic_1974, weihs_energetic_1980, videler_energetic_1982} who predicted the existence of a transition regime, $\operatorname{Re}_{\bar{U}}\in[20,200]$, in which burst coast swimming becomes more efficient than continuous swimming.

As the Reynolds number increases further, the gap between the burst coast and the C-start escape patterns continues to grow. This transition can be attributed to the increased efficiency of the intermediate coasting phase. In fact, for $\operatorname{Re}_{\bar{U}}\ge 200$, the drag coefficient of a rigidly gliding fish can be up to 4 times smaller than that of an actively swimming fish \cite{lighthill_large-amplitude_1971, webb_swimming_1971, wu_kinematics_2007}. This is thought to drive changes in the swimming style of fish who, when growing in length, transition from the viscous hydrodynamic regime to the inertial hydrodynamic regime and replace their continuous swimming style with predominantly burst coast swimming \cite{weihs_energetic_1980}. This biological transition has been termed an \textit{adaptive energy sparing mechanism} \cite{weihs_energetic_1980}. For the high Reynolds number regime, our analysis supports this hypothesis, suggesting that burst-coast swimming patterns convert limited energy into greater overall distance compared to patterns involving continuous swimming. Further study  to corroborate this hypothesis is needed since the model used is two-dimensional, and fish change their body form during growth \cite{voesenek_biomechanics_2018}.

\section{Generalizing escape patterns across energy budgets}
\label{sec:energy-budgets}
\begin{figure}[thbp]
\begin{minipage}[t]{0.03\textwidth}
(a)
\end{minipage}\hfill
\begin{minipage}[t]{0.30\textwidth}
\includegraphics[width=\textwidth, valign=t]{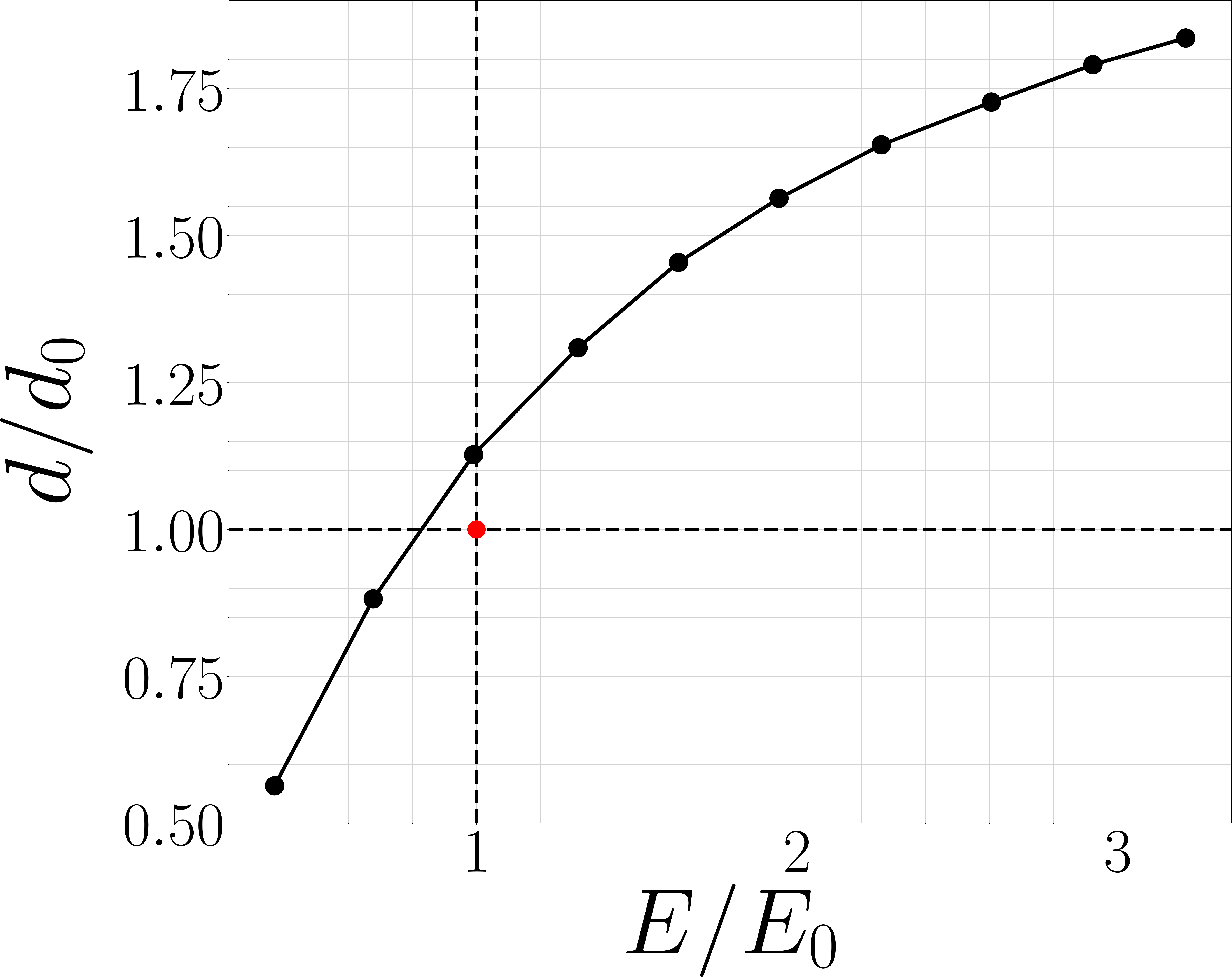}
\end{minipage}\hfill
\begin{minipage}[t]{0.03\textwidth}
(b)
\end{minipage}\hfill
\begin{minipage}[t]{0.30\textwidth}
\includegraphics[width=1.0\textwidth, valign=t]{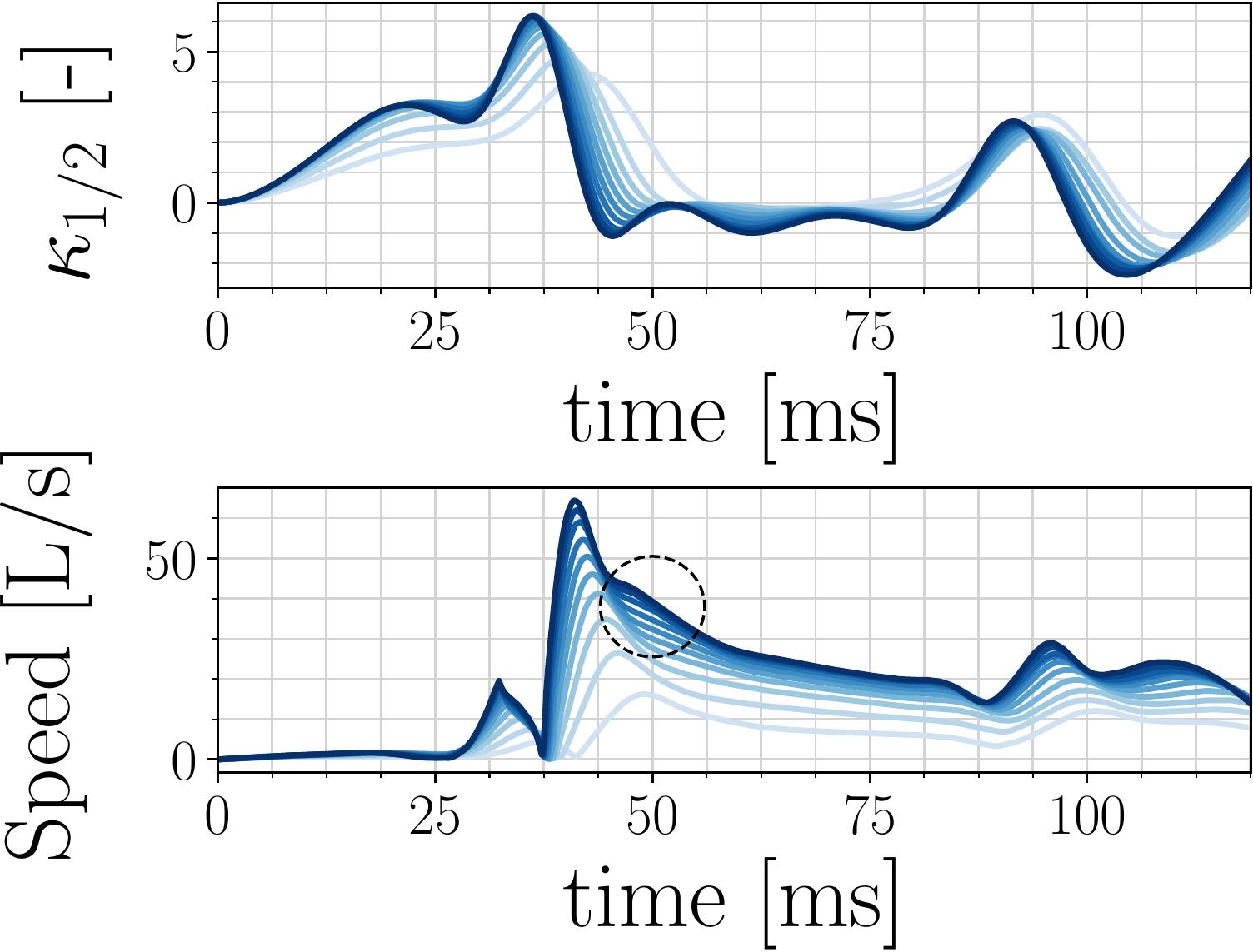}
\end{minipage}\hfill
\begin{minipage}[t]{0.03\textwidth}
(c)
\end{minipage}\hfill
\begin{minipage}[t]{0.30\textwidth}
\includegraphics[width=0.79\textwidth, valign=t]{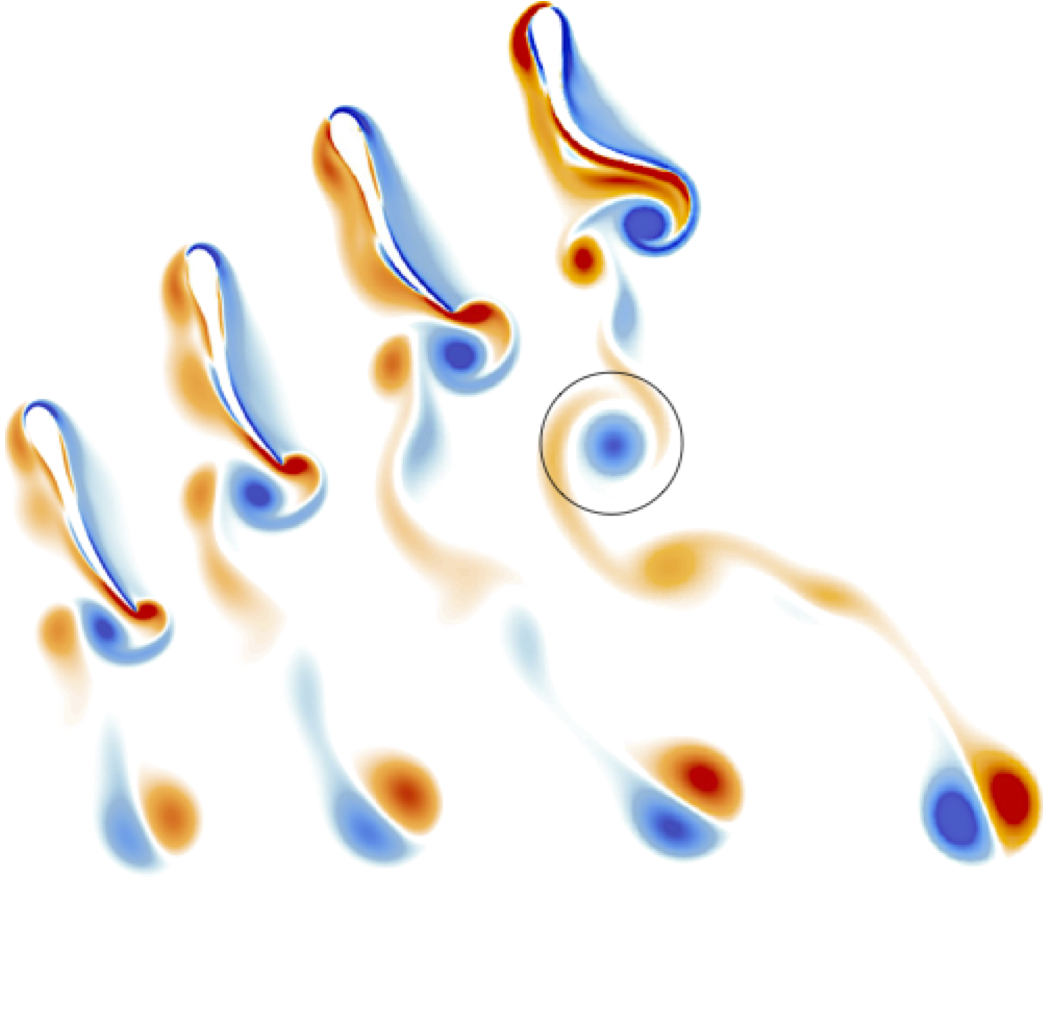}
\end{minipage}
\caption{Influence of energy budget on escape distance and learned strategy. (a) Normalized escape distance as a function of normalized energy expenditure for 10 escapes with energy budgets linearly spaced between $\frac{1}{3} E_0$ and $3 E_0$. (b) Swimmer midpoint curvature $\kappa_{1/2}$, and swimmer speed as a function of time for the 10 escapes. The higher energy escapes are represented by dark blue whilst the lower energy escapes are represented with light blue. The lightest blue curve corresponds to an energy budget $\frac{1}{3} E_0$, and the darkest blue corresponds to an energy budget $3 E_0$. All quantities are plotted for escapes of duration $118.8 \si{\milli\second}$ and are simulated using the same RL policy (for training details see Appendix). (c) Vorticity fields in the wake of four escapes of energy budgets, from left to right, $\frac{1}{3} E_0$, $\frac{9}{10} E_0$,  $\frac{6}{5} E_0$, and $3 E_0$. Snapshots of the vorticity field are taken at $t=75\si{\milli\second}$ for all 4 escapes. The dotted, circled region denotes a secondary vortical structure which emerges at higher energy budgets.}
\label{img:strategy-across-energy-budgets}
\end{figure}

The RL swimmer was endowed with different energy budgets and the resulting escape patterns were visualized in Figure \ref{img:strategy-across-energy-budgets}(c). We found that the learned swimmer was able to modulate its motions in order to travel further at higher energies (see Figure \ref{img:strategy-across-energy-budgets}(a)). Which aspects of the escape strategy does it modify to achieve this? Plots of the midpoint curvature for escapes of different energies (Figure \ref{img:strategy-across-energy-budgets}(b)) evidence that, as the energy budget increases, the swimmer performs more pronounced, faster C-bends. As a result, the maximum speed is attained at a higher value earlier in the escape, and the overall escape distance increases. This can be seen by the peak midpoint curvature, as well as the peak speed, moving upwards and to the left as the energy budget increases (Figure \ref{img:strategy-across-energy-budgets}(b)).

This strategy can be interpreted in terms of the Duty Cycle defined as $DC = T_{\textrm{burst}}/T_{\textrm{total}}$, where $T_{\textrm{burst}}$ is the time spent accelerating before the coasting phase, and $T_{\textrm{total}}$ is the total time of the escape bout. In this case, $T_{\textrm{burst}}$ corresponds approximately to the time where the peak speed is attained. As can be seen in Figure \ref{img:strategy-across-energy-budgets} (b), the peak speed is attained earlier for higher energies so the duty cycle decreases as the energy budget increases. In contrast, experimental data on the duty cycle of fish swimming at cruising speeds indicates that fish increase their duty cycle when higher speeds are required \cite{li_burst-and-coast_2021}. This suggests, that for escapes where high acceleration is required in order to increase escape distance in a fixed amount of time, the Duty Cycle of burst-coast strategies should be made as small as possible.

Another strategy learned by the swimmer consists in using the energy leftover from the C-bend and swim phases in order to perform slight undulations during the coasting phase. These undulations leave an imprint on the escape patterns in the form of a secondary vortical structure (circled region in Figure \ref{img:strategy-across-energy-budgets}(c)). However, this secondary vortical structure is absent from escapes of lower energy budget. This indicates that undulatory motions during the coasting phase are second-order to increasing escape distance, when compared to forming more pronounced C-bends or swimming more energetically. Indeed, for high energy escapes which include undulations during the coast phase ($E_{\textrm{budget}}\ge E_0$), the swimmer already performs C-bends very close to geometric/mechanical limit $\kappa=2\pi$ and undulates with curvature close to $\kappa = 2.5$ in the final swim phase. Thus, not being able to perform a C-bend past $\kappa=2\pi$ due to mechanical constraints, and swim with curvature past $\kappa = 2.5$ during the final phase because of increased drag, the leftover energy is used to undulate during the coast phase. This extends the high-speed phase of the escape (circled in Figure \ref{img:strategy-across-energy-budgets}(b)) and improves the overall distance. The saturation of the C-bend to the max possible curvature $\kappa=2\pi$ and the swim phase to $\kappa = 2.5$ are possible factors causing the rate of conversion of energy to distance to flatten off as the energy budget increases (Figure~\ref{img:strategy-across-energy-budgets}). 

\section{Propulsive mechanisms}

\begin{figure}[htbp]
\centering
\begin{minipage}[t]{0.01\textwidth}
(a)
\end{minipage}
\begin{minipage}[t]{0.47\textwidth}
\includegraphics[width=0.8\textwidth, valign=t]{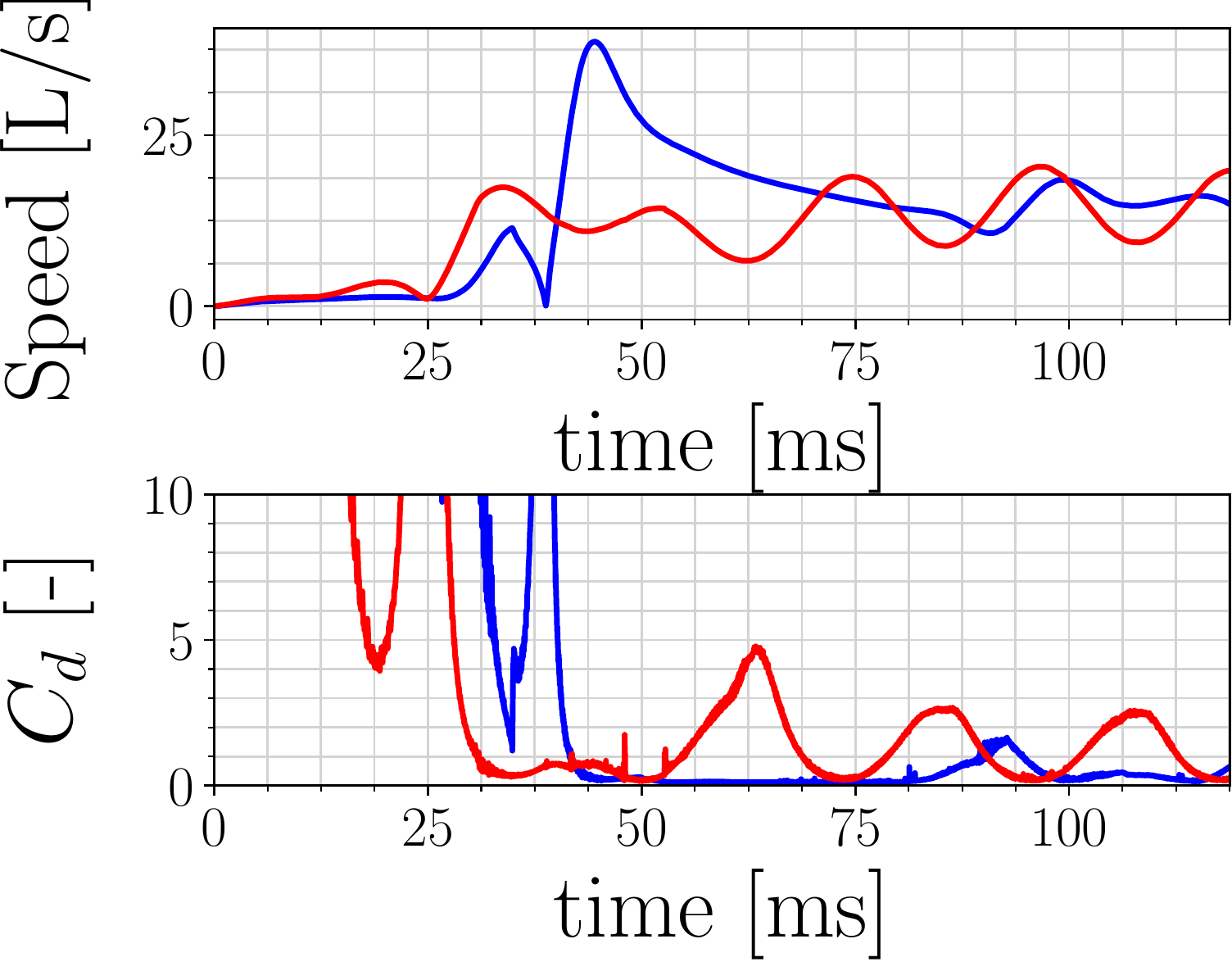}
\end{minipage}
\begin{minipage}[t]{0.03\textwidth}
(b)
\end{minipage}
\begin{minipage}[t]{0.47\textwidth}
\includegraphics[width=0.8\textwidth, valign=t]{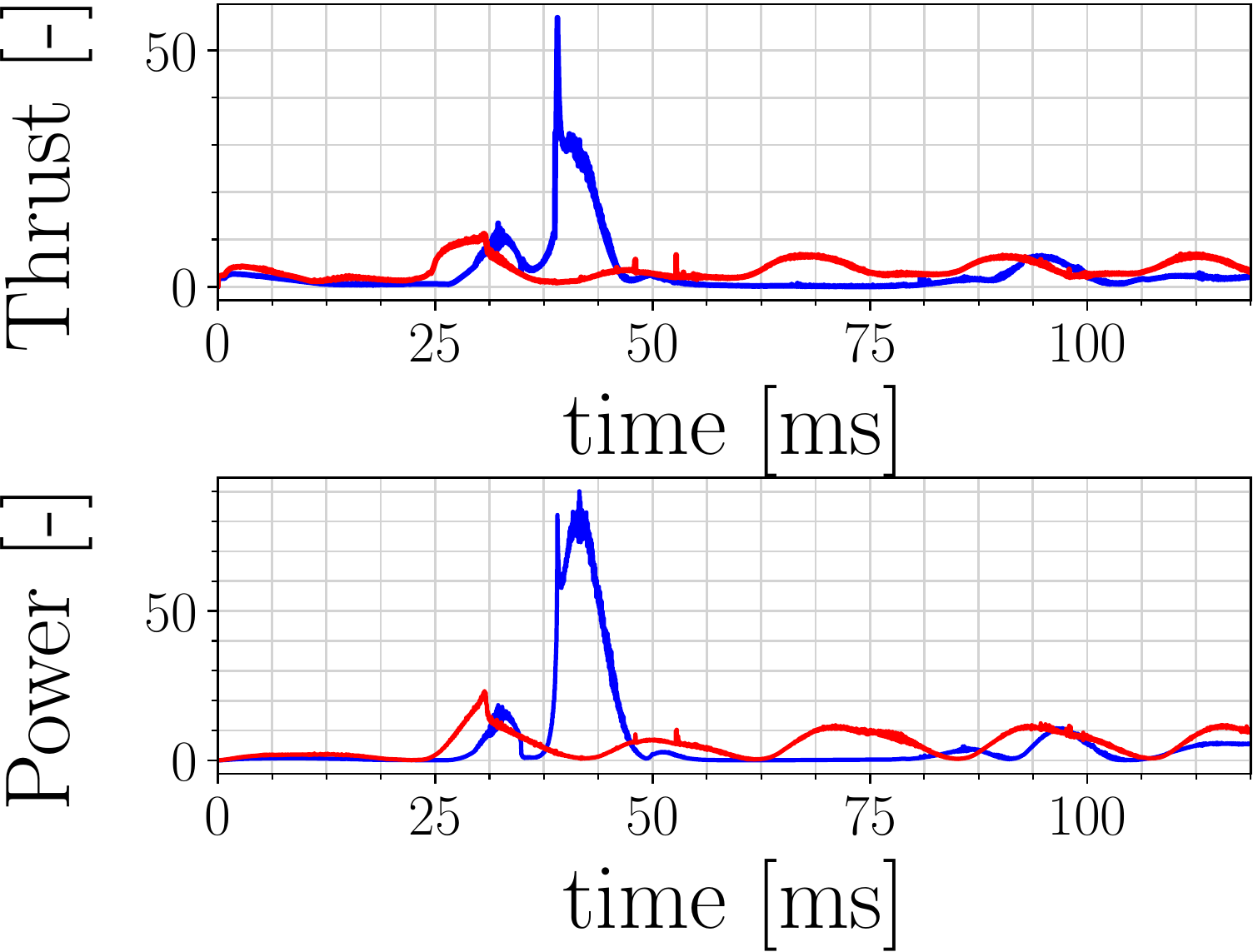}
\end{minipage}\hfill

\caption{Escape speed, drag coefficient, thrust force, and power consumption \BLUE{for equal-energy learned burst-coast pattern vs. optimized C-start pattern. The energy expended by the RL escape (blue) and C-start escape (red) is in both cases equal to $E_0$.} (a) Speed and drag coefficient of the burst-coast pattern learned through RL (blue) vs. the C-start pattern (red). The speed is given in swimmer lengths per second ($L/s$) and the drag coefficient is defined as $C_d$ = $\frac{F_d}{\rho U^2 L/2}$, where $F_d$ is the drag force experienced by the swimmer, $\rho$ is the density of water, $U$ is the swimmer speed relative to the fixed lab reference frame, and $L$ is the swimmer length. (b) Thrust force and deformation power of the burst-coast pattern (blue) vs. the C-start pattern (red). For details on the computation and non-dimensionalization of the drag force $F_d$, thrust force, and deformation power see Appendix A.}
\label{img:propulsive-forces}
\end{figure}

The escape speed, drag coefficient, thrust, and power were computed for the RL burst-coast pattern and compared to the C-start escape pattern with the optimal parameters from \cite{gazzola_c-start_2012} in Figure \ref{img:propulsive-forces}. The RL swimmer produces a significantly higher peak thrust force than the C-start during the starting phase of the escape (Figure \ref{img:propulsive-forces} (b)). As alluded to in Section \ref{sec:learned-vs-optimized}, this is a result of the RL swimmer curling its body more than the C-start swimmer during the initial C-bend motion. As a result, a higher peak power and maximum speed are attained (Figures \ref{img:propulsive-forces} (a, b)). Moreover, the RL swimmer achieves a $13.5g$ peak acceleration, significantly greater than the $9.4g$ peak acceleration created by the lower thrust C-start swimmer.

Furthermore, the RL swimmer produces close to zero thrust after the initial C-bend whilst the C-start swimmer continually produces thrust and expends power during the escape as a result of the continuous swimming pattern (Figure \ref{img:propulsive-forces}(b)). Thus, the learned escape pattern uses more energy during the initial propulsion but saves energy by coasting in the remaining time. Why does this result in increased escape distance using the same energy ? When plotting the drag coefficient $C_d$ against time for the two escape patterns we observe that the drag during the coasting phase of the escape is close to zero for the RL swimmer whilst the C-start swimmer continually experiences elevated drag due to the swimming motion. This analysis suggests that expending most energy in a strong initial C-bend and then coasting is advantageous due to the decrease of fluid dynamic drag when compared to swimming.

\begin{figure}[thbp]
\begin{minipage}[t]{0.03\textwidth}
(a)
\end{minipage}\hfill
\begin{minipage}[t]{0.36\textwidth}
\includegraphics[width=\textwidth, valign=t]{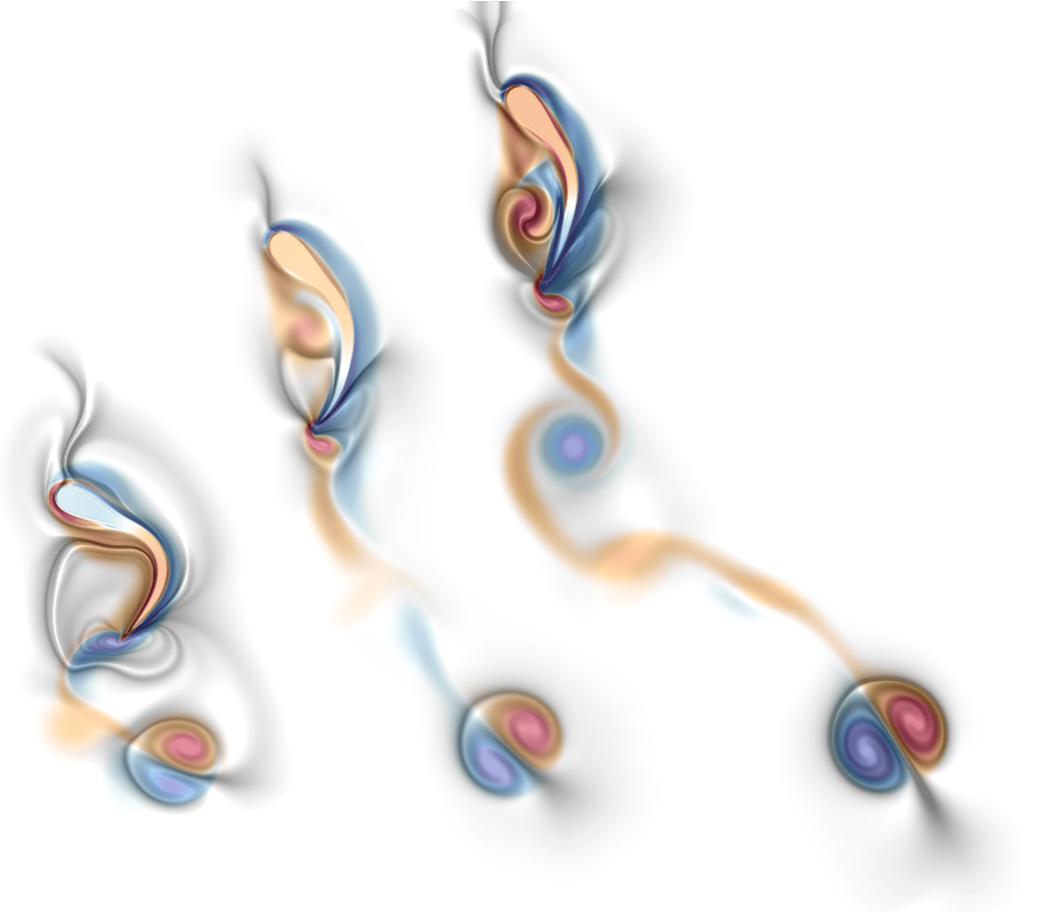}
\end{minipage}\hfill
\begin{minipage}[t]{0.03\textwidth}
(b)
\end{minipage}\hfill
\begin{minipage}[t]{0.58\textwidth}
\includegraphics[width=\textwidth, valign=t]{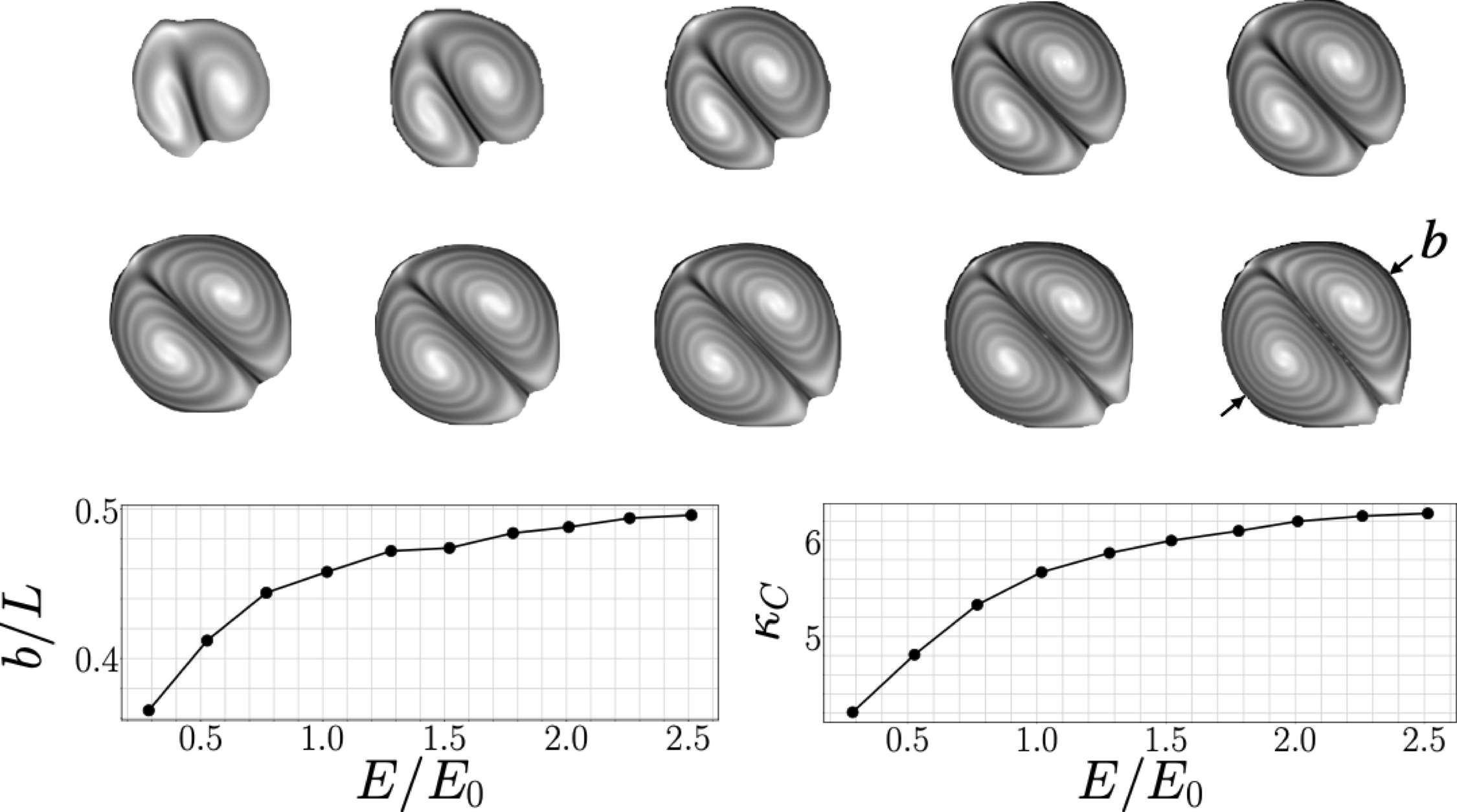}
\end{minipage}%
\caption{Characterization of Lagrangian Coherent Structures at different energy budgets. (left) Vorticity fields overlayed onto the Finite Time Lyapunov Exponent Field (FTLE) of three different escapes. From left to right are the C-start, the artificial swimmer with energy budget $E_0$, and the artificial swimmer with energy budget $3 E_0$. The FTLE is visualized on a spectrum from white to black. The darker colors represent a higher value of the FTLE. The LCS are the 'ridges' of the FTLE field, i.e. the darkest contours on the visualization. All FTLE fields are computed with the same integration length and displayed at the same time instant (see Appendix). (a) A visualization of the lump of fluid ejected by the swimmer, the average vortex diameter $b$, and the peak C-bend midpoint curvature $\kappa_C$ for escapes with energy budgets linearly spaced between $\frac{1}{3}E_0$ and $3 E_0$, shown to scale. The lumps of fluid are extracted by localizing the high ridge (LCS) values in the respective FTLE field. The average vortex diameter $b$ is normalized by the swimmer length $L$ as a function of the energy budget.}
\label{img:vortex-balls}
\end{figure}

Finally, the hydrodynamic mechanisms exploited by the artificial swimmer were analyzed using Lagrangian Coherent Structures (LCS) \cite{haller_lagrangian_2000}. Well defined LCS are characterized by negligible flux across their surface, acting as transport barriers within the flow \cite{shadden_definition_2005}. In Figure \ref{img:vortex-balls} (a) the resulting LCS are superimposed onto the vorticity fields for the C-start and two RL escapes of different energy. The LCS evidence one predominant coherent structure: a lump of fluid carried by the counter-rotating vortex pair. This has previously been observed in \cite{huhn_quantitative_2015}, and is a common structure found in fish escape sequences.

Furthermore, we notice that the lump of fluid ejected by the C-start escape motion (left Figure \ref{img:vortex-balls} (a)) is less symmetric, and smaller than that of the RL escape (middle Figure \ref{img:vortex-balls} (a)). Thus, the RL swimmer uses more of the allocated energy budget to trap and accelerate a large volume of water opposite to the escape direction, whilst the C-start swimmer uses less energy to propel the initial ball of water, but compensates by using the leftover energy to swim continuously for the rest of the escape. Moreover in the case of RL, the vortex dipole and its lump of fluid are more aligned with the direction of the motion than what is observed in C-start. This continuous alignment minimizes the time derivative of the vorticity linear impulse  and as such the drag experienced by the swimmer \cite{koumoutsakos_high-resolution_1995}. Finally, we found that as the energy budget increased, the average diameter of the lump of fluid ejected by the swimmer began to flatten off close to $b\approx\frac{1}{2} L$ (see Figure \ref{img:vortex-balls} (b)). Since the peak curvature of the C-bend cannot increase beyond $2\pi$ ($\kappa_C\le 2\pi$), the swimmer cannot convert higher energy into more escape distance indefinitely by forming more pronounced C-bends (Figure \ref{img:strategy-across-energy-budgets} (b)), thus limiting the maximum escape distance attainable at a given Reynolds number and morphology. 

\section {Conclusions}

This study explores the use of deep reinforcement learning to discover swimming escape patterns which maximize distance given a fixed amount of energy. In contrast to an optimization process with an overarching goal, RL explores an array of incremental processes allowing it to learn a range of escape patterns. Our results indicate that maximum swimming distance can be achieved through short bursts of accelerating motion interlinked by phases of powerless gliding. In the context of fish escaping from \BLUE{disturbances}, we find that, at higher Reynolds numbers, burst-coast escape patterns result in greater escape distances than burst-swim escape patterns (C-starts), but C-starts propel the swimmer away from the initial position faster. \BLUE{This suggests that, larval zebrafish performing C-starts may not only be aiming to maximize escape distance, but their internal reward function may further include a notion of ‘urgency’ to distance themselves from their initial position as quickly as possible. Future studies may benefit from encoding urgency into the reward function and observing how the resulting swimming escape patterns change.}

Contrary to an optimization setting which requires using domain-specific knowledge to pre-define stages of motion, the reinforcement learning setting is free from prior bias on the functional form of the escape pattern. This additional freedom results in escape patterns that outperform those obtained by optimization, for the same energy budget. Moreover, we find that training the swimmer to control its motions as a function of the energy budget produces a ``kaleidoscope" of escape patterns that reveal practical flow optimization principles for efficient swimming. Studying the learned strategies indicates that the formation of a C-shape, a coasting phase, and a final swim motion are necessary components of distance maximizing escapes across energy budgets. Other strategies, such as slight undulations during the coasting phase, are eliminated as the energy budget decreases, indicating their second-order, but non-negligible, importance for achieving rapid propulsion from rest. This suggests that reinforcement learning can more robustly discover swimming escape patterns than methods based on reverse engineering via optimization.

Finally, we emphasize that RL is a data efficient learning methodology. In the current problem setup, each action is determined by 9 real valued parameters, and each escape consists of around 7 actions. If this problem were to be solved with a stochastic optimizer, this amounts to solving a $n=63$ dimensional constrained optimisation problem. Stochastic optimizers like CMA-ES find global optima for a variety of functions using $300n-500n^2$ function evaluations \footnote{N. Hansen, S. Kern, Evaluating the CMA Evolution Strategy on Multimodal Test Functions, Parallel Problem Solving from Nature - PPSN VIII, 2004.}. Thus, the computational cost can be anywhere between $18'900-1'984'500$ simulations.  Furthermore, the RL policy can be controlled according to a real-valued energy budget, producing a wide array of swimming escapes. Approximating this level of fine-grained control with an optimisation approach would require repeating the optimisation for many different energy budgets (e.g. $N$ times) - further increasing the computational cost to anywhere between $18'900N-1'984'500N$ simulations. Compounded with the fact that the optimal number of actions to be taken given the energy and time constraints \emph{is not known in advance} and should be learned (as done by reinforcement learning), renders the problem totally impractical if approached with optimization-based reverse engineering. On the contrary, we find that solving the full problem via RL is tractable using only around 150'000 simulations. 

In summary we find that  RL is a powerful tool for the discovery of swimming escape patterns under energy constraints. The identified escape patterns deepen our understanding of the hydrodynamic mechanisms that are exploited by natural swimmers.

\appendix

\section{NUMERICAL METHOD}

We use a simplified 2D geometric model of a 5 days post-fertilization zebrafish ~\cite{gazzola_c-start_2012}. The swimmer shape is described by the body half-width $w(s)\in\mathbb{R}$ along the curvilinear coordinate $s\in[0,L]$ for a body length $L\in\mathbb{R}$.
\begin{equation}
\centering
    w(s)=
    \begin{cases}
    w_{h} \sqrt{1-\left(\frac{s_{b}-s}{s_{b}}\right)^{2}}& 0\leq s<s_{b}\\
    \left(-2\Delta w-w_{t}\Delta s\right)\delta s_b^{3} \\
    + \left(3\Delta w+w_{t}\Delta s\right)\delta s_b^{2}& s_b\leq s<s_{t}\\
    +w_{h}\\
    w_{t}-w_{t}\left(\frac{s-s_{t}}{L-s_{t}}\right)^{2}& s_t\leq s<L
    \end{cases}\,.
\end{equation}
Above, $\Delta w=w_{t}-w_{h}$, $\Delta s=s_{t}-s_{b}$ and $\delta s_b=\frac{s-s_{b}}{\Delta s}$, where $s_{b}=0.0862 L, s_{t}=0.3448 L, w_{h}=0.0635 L, w_{t}=0.0254 L$. In order to resolve the head and the tail of the swimmer the spacing along the midline in the first and last $10\%$ of the body is linearly increased from $\Delta x/8$ to $\Delta x/\sqrt{2}$, where $\Delta x$ denotes the uniform resolution of the computational grid.

The flow field generated by the motion of the artificial swimmer is simulated by solving the two-dimensional, incompressible Navier-Stokes equations with volume penalisation~\cite{Angot1999,Coquerelle2008,Gazzola2011}:
\begin{equation}
    \begin{split}
\nabla \cdot \boldsymbol{u} &=0\,, \\
\frac{\partial \boldsymbol{u}}{\partial t}+\boldsymbol{u} \cdot \nabla \boldsymbol{u} &=-\frac{\nabla p}{\rho}+\nu \nabla^{2} \boldsymbol{u}+\lambda \chi\left(\boldsymbol{u}_{s}-\boldsymbol{u}\right)\,.
\end{split}
\label{eq:nse}
\end{equation}
Here, $\boldsymbol{u}(x,t)\in\mathbb{R}^2$ corresponds to the fluid velocity and $p(\boldsymbol{x},t)\in\mathbb{R}$ to the pressure. The fluid properties are determined by the viscosity $\nu\in\mathbb{R}$ and the fluid density $\rho\in\mathbb{R}$. The fluid-structure interaction is modeled by the penalty term $\lambda \chi\left(\boldsymbol{u}_{s}-\boldsymbol{u}\right)$, where $\boldsymbol{u}_{s}\in\mathbb{R}^2$ denotes the combined translational, rotational, and deformation velocity of the swimmer. The characteristic function $\chi(\boldsymbol{x},t)\in\mathbb{R}$ is $1$ inside the swimmer, $0$ elsewhere. The equation is solved in a two-dimensional domain $\boldsymbol{x}\in\Omega\subset\mathbb{R}^2$, over a time-interval $t\in[0,T]\subset\mathbb{R}$. The domain was chosen to be four times the length of the swimmer $\Omega=\left[4 L, 4 L\right]\subseteq\mathbb{R}^2$ and we ran the solver up to time $T_{\text{max}}=118.8$ ms.

In order to solve the equation (\ref{eq:nse}) we discretized the equation on a uniform grid in space. On this grid, the spatial derivatives were approximated using second-order centered finite differences. For this purpose we used the uniform grid library Cubism~\cite{rossinelli_11_2013}. The time-stepping is performed using explicit Euler, where the time-step $\Delta t$ was adopted such that the CFL number was constrained to $0.1$. To ensure momentum conservation and stability the penalty parameter is set to $\lambda=1/\Delta t$~\cite{Coquerelle2008}. For the characteristic function we use a second order approximation of the Heaviside function~\cite{Towers2009}.  During the reinforcement learning, we use $512\times 512$ gridpoints. All quantities of interest are subsequently computed at the higher resolution of $1024\times 1024$ gridpoints.

In the following we describe the operator splitting formalism used to compute the time-step. Starting with the velocity field $\boldsymbol{u}^t$ at timestep $t$ we computed an intermediate velocity $\boldsymbol{u}^*$ by performing advection and diffusion
\begin{equation}
    \boldsymbol{u}^*=\boldsymbol{u}^t+\Delta t(\nu \nabla^{2} \boldsymbol{u}-\boldsymbol{u} \cdot \nabla \boldsymbol{u})\,.
\end{equation}
The resulting velocity field is non-divergence-free and we used pressure projection~\cite{Chorin1968}
\begin{equation}\label{eq:pressure-projection}
    \boldsymbol{u}^{**}=\boldsymbol{u}^{*}-\Delta t\frac{\nabla p^{t+1}}{\rho}\,.
\end{equation}
The pressure field was computed by solving the Poisson equation that results when taking the divergence of equation (\ref{eq:pressure-projection}) and using that the obstacle velocities can be non-divergence free due to the deformation component $\nabla\cdot\boldsymbol{u}^{**}=\chi\nabla\cdot\boldsymbol{u}_{s}^{t+1}$
\begin{equation}
    \Delta p^{t+1}=\frac{\rho}{\Delta t}\left[\nabla\cdot\boldsymbol{u}^*-\sum\limits_{s=1}^{N_s}\chi\nabla\cdot\boldsymbol{u}_{s}^{t+1}\right]\,.
\end{equation}
We conclude the time-step by employing the penalisation force on the field
\begin{equation}\label{eq:penalisation}
  	\boldsymbol{u}^{t+1}=\boldsymbol{u}^{**}+\chi(\boldsymbol{u}_s^{t+1}-\boldsymbol{u}^{**})\,,
\end{equation}
where we used $\lambda=1/\Delta t$.

Using the numerical solution of the Navier-Stokes equation we can compute the work done by the deformation of the swimmer body $\boldsymbol{u}_{\textrm{def}}$ on the surrounding flow, which can be thought of as the muscle input power, as:
\begin{equation}
E(t) = \int_0^t\left[{\int_{\partial\Sigma}\boldsymbol{u}_{\textrm{def}}\cdot d\boldsymbol{F}}\right]\mathrm{d} t\,.
\label{eq:energy-fish-on-flow}
\end{equation}
\noindent Note that the computational model used does not account for muscle dynamics so the muscle input power cannot directly be computed, only approximated. 

The energy values reported are non-dimensionalized by $M L^2 / T_{\textrm{prop}}^2$, where $M$ is the swimmer mass, $L$ the swimmer length, and $T_{\textrm{prop}}$ is the propulsive swimming period. The deformation power follows from \ref{eq:energy-fish-on-flow}, computed as: $P_{\textrm{def}} = \frac{d E(t)}{d t}$ and is non-dimensionalized by $M L^2/T_{\textrm{prop}}^3$. The propulsive thrust $F_t$ and drag force $F_d$ are computed as displayed in equations \ref{eq:thrust} and \ref{eq:drag} are non-dimensionalized by $M L^2/T_{\textrm{prop}}^2$.

\begin{equation}
F_t = \int_{\partial\Sigma}\left(\boldsymbol{u}\cdot d\boldsymbol{F} + \left|\boldsymbol{u}\cdot d\boldsymbol{F}\right|\right) / \left(2\left|\boldsymbol{u}\right|\right)
\label{eq:thrust}
\end{equation}

\begin{equation}
F_d = \int_{\partial\Sigma}\left(\boldsymbol{u}\cdot d\boldsymbol{F} - \left|\boldsymbol{u}\cdot d\boldsymbol{F}\right|\right) / \left(2\left|\boldsymbol{u}\right|\right)
\label{eq:drag}
\end{equation}

In \ref{eq:energy-fish-on-flow}, \ref{eq:thrust}, \ref{eq:drag}, $\partial\Sigma$ denotes the swimmer surface, and $\mathrm{d}\boldsymbol{F}$ is the force acting on the swimmer comprised of viscous and pressure-based forces $\mathrm{d}\boldsymbol{F} = \mathrm{d}\boldsymbol{F}_{P} + \mathrm{d}\boldsymbol{F}_{\nu} = 2\mu\boldsymbol{D}\cdot\boldsymbol{n}\mathrm{d} S - P\boldsymbol{n}\mathrm{d} S$. Here, $\boldsymbol{D}=\frac{1}{2}(\nabla\boldsymbol{u} + \nabla\boldsymbol{u}^T)$ is the strain-rate tensor, $P$ is the surface pressure, $\mu$ is the dynamic viscosity,  $\boldsymbol{n}$ is the surface normal, and $\mathrm{d}S$ is the infinitesimal surface element.

\section{REINFORCEMENT LEARNING}
During the escape, the swimmer senses its cylindrical coordinates $(d, \phi)\in\mathbb{R}^2$, its center-of-mass velocity $\boldsymbol{v}\in\mathbb{R}^2$, as well as its orientation and angular velocity $\theta, \dot{\theta}\in\mathbb{R}$ relative to the fixed laboratory frame. Furthermore, it has access to the remaining energy available for the escape $E_{\textrm{to-go}}\in\mathbb{R}$, and a memory of its action from two previous time steps $\boldsymbol{a}_t, \boldsymbol{a}_{t-1}$. These perceptive abilities form the state vector $\boldsymbol{s} = (d, \phi, \boldsymbol{v}, \theta, \dot{\theta}, E_{\textrm{to-go}}, \boldsymbol{a}_t, \boldsymbol{a}_{t-1})\in\mathbb{R}^{25}$. The energy available for the escape, $E_{\textrm{to-go}}\in\mathbb{R}$, is computed as the difference between the available energy $E_{\text{budget}}$ and the work done by the swimmer on the surrounding flow.

Given each current state, the swimmer is able to influence its mid-line configuration by scheduling changes in curvature. In particular it can select the amount of mono-lateral (baseline) curvature $B(s,t)\in\mathbb{R}$, undulatory curvature $K(s,t)\in\mathbb{R}$, the travelling wave phase $\tau_L\in\mathbb{R}$, the overall phase $\phi\in\mathbb{R}$, and the duration of each transition $\Delta t\in\mathbb{R}$. Since the baseline and undulatory curvature are each parameterized by 6 control points, of which 3 are free, the actions are given as $    \boldsymbol{a}=(B_1, B_2, B_3, K_1, K_2, K_3, \tau_{L}, \phi, \Delta t)\in\mathbb{R}^9$.
% %
% \begin{equation}
%     \boldsymbol{a}=(B_1, B_2, B_3, K_1, K_2, K_3, \tau_{L}, \phi, \Delta t)\in\mathbb{R}^9\,.
% \end{equation}
% %
Here, $B_i,K_i\in\mathbb{R}$ for $i=1,2,3$ denote the 3 controllable baseline and 3 controllable undulatory curvatures on the swimmer mid-line. As per experimental observations of zebrafish performing fast-starts~\cite{muller_flow_2008}, we constrain the maximum curvature of the artificial swimmer to $|\kappa(s,t)|\le 2\pi/L$ and the action duration to $\Delta t \in [0.5 T_{\textrm{prop}}, T_{\textrm{prop}}]$. The phases are constrained to $\tau_L, \phi_L \in\left[0, 2\pi\right]$. The transition between actions takes place in time by cubic-interpolation with derivatives equal on both sides of extrema to ensure continuity.

The swimmer starts in a zero curvature configuration at the center of a square simulation domain and is assigned a random energy budget $E_{\textrm{budget}}$ sampled uniformly between $\frac{1}{3}E_0$ and $3 E_0$. The episode terminates if the swimmer has depleted its energy budget, the allocated time is surpassed, or spatial constraints are violated. The maximum time for the escape is set to $T_{\textrm{max}} = 118\ \si{\milli\second}$, equal to the time-length of the C-start escape \cite{gazzola_c-start_2012}. Furthermore, the episode is terminated and the swimmer is penalized with $r=-10$ if the swimmer changes orientation by $|\Delta\theta|\ge\pi/2$. Since a typical zebrafish only changes its orientation by approximately $44$ degrees during a fast-start \cite{muller_flow_2008}, this constraint restricts the exploration space, but is sufficiently relaxed to not significantly influence the escape pattern. To model the scenario of a predator approaching from behind the zebrafish, we enforce that the swimmer center-of-mass remains in an infinite triangular region starting $0.2L$ behind the swimmer tail endpoint and with $40$ degrees aperture. If the swimmer exits the allowable region the episode is terminated and the swimmer is penalized with $r=-10$.

The off-policy actor-critic reinforcement learning algorithm V-RACER \cite{novati_remember_2019} implemented in smarties (see \url{https://github.com/cselab/smarties}) is employed for 1'000'000 state-action-reward observations. The neural network used to approximate the policy network and state value function has 3 hidden layers of 32 parameters each. A discount factor of $\gamma=0.99$ is used, the batch size is set to $128$, the exploration noise probability is set to $0.2$, and the learning rate for stochastic gradient descent is set to $0.0001$. The other hyper-parameters are left as described in the original publication.

\section{LAGRANGIAN COHERENT STRUCTURES}

The Lagrangian Coherent Structures (LCS) are the ridges of the Finite Time Lyapunov Exponent Field (FTLE). The FTLE is a scalar field $\sigma(\boldsymbol{X}_0)$ which characterizes the amount of stretching about the trajectory of a passive flow particle, from the location $\boldsymbol{X}_0$ to the location $\boldsymbol{X}$ during time $T$ \cite{conti_gpu_2012}. The trajectory ${\boldsymbol{X}}(t)$ of a particle located at position $\boldsymbol{X}_0={\boldsymbol{X}}(t_0)$ at time $t_0$ can be obtained by integrating
\begin{equation}
\frac{\mathrm{d}}{\mathrm{d}t}{\boldsymbol{X}}\left(t\right) =\boldsymbol{u}\left[\boldsymbol{X}\left(t\right), t\right]\,,
\end{equation}

\noindent where $\boldsymbol{u}$ is the flow velocity field. By following the particle trajectories for a time length $T$, we obtain the particle flow map which gives the particle position at later times $\boldsymbol{X}_0\rightarrow \boldsymbol{X}(t_0+T)=\phi(\boldsymbol{X}_0,T)$. From the flow map we can obtain the (right) Cauchy-Green deformation tensor $\Delta$, which quantifies the stretching of an infinitesimal material line
\begin{equation}
    \Delta(\boldsymbol{X}_0)=\left[\frac{\partial\phi(\boldsymbol{X}_0,T)}{\partial \boldsymbol{X}_0}\right]^\top \frac{\partial \phi(\boldsymbol{X}_0,T)}{\partial\boldsymbol{X}_0}
\end{equation}
The Finite-Time Lyapunov Exponent (FTLE) is then defined as the square root of the logarithm of the maximum eigenvalue $\lambda_{\text{max}}$ of the Cauchy-Green deformation tensor $\Delta$ normalized by the integration time $T$
\begin{equation}
    \sigma(\boldsymbol{X}_0)=\frac{1}{T}\sqrt{\log(\lambda_{\max}(\Delta(\boldsymbol{X}_0))}
\end{equation}
The FTLE fields reported in this study are calculated by integrating in forward-time using the open source software $\texttt{FTLE2D}$ \cite{conti_gpu_2012}. We use a set of 158 velocity fields of the $118.8\si{\milli\second}$ escape, spaced at equal time intervals. The FTLE is computed for the first 100 velocity fields, thus the receding integration time-horizon is 58 time steps. The time-horizon is chosen to be sufficiently long in order to adequately locate the LCS on the FTLE fields.

\bibliography{main}
\end{document}

%% file: caption-mod.tex
\makeatletter
\newcommand\Scaption[1]{%
  \sbox\@tempboxa{\small~#1}%
  \ifdim \wd\@tempboxa >\hsize
    \small~#1\par
  \else
    \global \@minipagefalse
    \hb@xt@\hsize{\hfil\box\@tempboxa\hfil}%
  \fi
  }
\makeatother